\documentstyle[12pt,cite]{article}

\def\Journal#1#2#3#4{{#1} {\bf #2}, #3 (#4)}

\def\NPB{{\em Nucl.Phys.}B}
\def\PLB{{\em Phys.Lett.}B}
\def\PRL{\em Phys.Rev.Lett.}
\def\PRD{{\em Phys.Rev.}D}
\def\ZPC{{\em Z.Phys.}C}

\def\SJNP{\em Sov.J.Nucl.Phys.}

\def\SPJETP{\em Sov.Phys.JETP}

\def\SPJETPL{\em Sov.Phys.JETP Lett.}

\def\ibid{\em ibid}

\def\GC{\em Grav\&Cosm}

\def\TMP{\em Sov.J.Theor.Math.Phys.}
\begin{document}

%\title{}

%\author{}

%\address{}


%\maketitle
%\abstracts{}

%\section{}
%\subsection{}

\begin{center}

{\bf Neutrino Oscillations in
Electromagnetic Fields}

{\bf A.M. Egorov, A.E. Lobanov,
A.I. Studenikin\footnote{\normalsize E-mail: studenik@srdlan.npi.msu.su}}

{\bf Department of Theoretical Physics, Moscow State University,

119899 Moscow,Russia}

%\vspace{0.5cm}

{\it Abstract}

\end{center}

{\it
Oscillations of neutrinos $\nu_L \leftrightarrow \nu_R$ in presence of
an arbitrary electromagnetic field are considered. We introduce the
Hamiltonian for the neutrino spin evolution equation that accounts for
possible effects of interaction of neutrino magnetic $\mu$ and
electric $\epsilon$ dipole moments with the transversal
(in respect to the neutrino momentum) and also the longitudinal
components of electromagnetic field.  Using this Hamiltonian we
predict the new types of resonances in the neutrino oscillations 
$\nu_L \leftrightarrow \nu_R$ in the presence of the field of an 
electromagnetic wave and in combination of an electromagnetic wave
 and constant magnetic field. The possible influence of the 
longitudinal magnetic field on neutrino oscillations is emphasized.}

 The electromagnetic properties of neutrinos are among the most
interesting issues in particle physics. Studies of the neutrino
electromagnetic properties could provide an important information about the
structure of theoretical model of particle interaction.
For instance, the discovery of the non-vanishing neutrino magnetic moment,
as well as the neutrino mass, would clearly
indicate that the Standard Model has to be generalized.

The non-vanishing
neutrino magnetic moment has also crucial consequences in
astrophysics.  As it has been shown in plenty of studies (see, for
example, \cite{Cis, Fu, SchV, VVOn, VVO, LM_Akh, LC_BM, Vol, Vidal_,
Smirnov_, APS, LS94, LS, ELS, LSg, KS, ALS, Barbieri}) that have
emerged during past decades, the neutrino conversions and
oscillations produced under the influence of transversal constant or
constant and twisting (in space) magnetic fields could be important
for evolution of astrophysical object, like the Sun and neutron
stars, or could result in sufficient effects while neutrinos
propagate through interstellar galactic media
\footnote{It should be noted here that the neutrino helicity flip
could be caused not only by the interaction with an external magnetic
field (or, as it is shown below with an electromagnetic wave) but
also by the scattering with charged fermions in the background (see,
for example, \cite{Ayala} and references therein)}.

In the previously performed studies of neutrino spin precession only
effects of the neutrino magnetic (or flavour transition) moment interaction
with transversal constant or twisting magnetic fields were considered
(see, for example, \cite{SchV} - \cite{LM_Akh},
\cite{Vol} - \cite{LSg}). The influence of the longitudinal component
of magnetic field is usually neglected because in the relativistic
limit it is suppressed. In the presence of magnetic fields the
neutrino evolution equation accounting for the magnetic moment
interaction can be received on the basis of relativistic wave
equations.  The usually discussed Hamiltonians for the neutrino time
evolution Schr$\ddot o$dinger equation can be derived by expanding
the exact Dirac Hamiltonian in powers of the neutrino kinetic
energy. In the lowest order there is no dependence on the
longitudinal to the neutrino momentum component of the magnetic
field, because the transversal components of the field acquire a
factor $\gamma=(1-\beta^2)^{-1/2}$ in the rest frame of neutrino
($\vec \beta=\vec v/c$, where $v$ is the neutrino speed).

The purpose of this paper is to generalize the Hamiltonian describing
the neutrino spin evolution for the case of the neutrino motion in an
arbitrary configuration of electromagnetic fields.
We derive \cite{LES99, hep-ph/9910476} the Hamiltonian that accounts
not only for the transversal to the neutrino momentum components of
electromagnetic field but also for the longitudinal components. With
the using of the proposed Hamiltonian it is possible to consider
neutrino spin precession in an arbitrary configuration of
electromagnetic fields including those that contain strong
longitudinal components. We also consider the new effect of the
neutrino spin precession that could appear when neutrinos propagate
in matter under the influence of a field of electromagnetic wave and
the superposition of electromagnetic wave and constant longitudinal
magnetic field.  The new types of resonances in the neutrino
oscillations $\nu_{L}\leftrightarrow\nu_{R}$ in such field
configurations are predicted. The influence of the longitudinal 
component of the magnetic field on the neutrino oscillations is 
discussed.

 The equation for the neutrino
spin evolution in electromagnetic field $F_{\mu \nu}$
is received on the basis of the Bargmann-Michel-Telegdi 
(BMT) equation \cite{BMT59} for the spin vector
$S^{\mu}$ of a neutral particle that has the following form
\begin{equation}
\begin{array}{c}
{dS^{\mu} \over d\tau} =2\mu \big\{ F^{\mu\nu}S_{\nu} -u^{\mu}(
u_{\nu}F^{\nu\lambda}S_{\lambda} ) \big\} +2\epsilon \big\{ {\tilde
F}^{\mu\nu}S_{\nu} -u^{\mu}(u_{\nu}{\tilde
F}^{\nu\lambda}S_{\lambda}) \big\}.
\end{array}
\label{A1}
\end{equation}
We suppose that the particle is moving with constant speed,
$\vec \beta=const$, in presence of an electromagnetic
field $F_{\mu\nu}$.
Here $\mu$ is the fermion magnetic moment and
${\tilde F}_{\mu\nu}$ is the dual electromagnetic field tensor.
Equation (\ref{A1}) covers also the case of a neutral fermion having
static non-vanishing electric dipole moment, $\epsilon$.  Note that
the term proportional to $\epsilon$ violates $T$ invariance.

Let us underline that eq.(\ref{A1}) accounts
for the direct interaction of a neutral fermion with
electromagnetic field $F_{\mu\nu}$. It should be noted here
that there could be an indirect influence of electromagnetic field on
a neutral fermion via one-loop finite-density contributions to the
particle self-energy in electromagnetic field.

The BMT equation (\ref{A1}) is derived in the frame of
electrodynamics. However, neutrino participates also in weak
interaction in which, contrary to the electromagnetic interaction,
$P$ invariance is not conserved.
Clearly, this fact has to be reflected in the form of equation that
describes the neutrino spin evolution in an electromagnetic field.
Our goal is to modify eq.(\ref{A1}) and to derive the new one which 
is appropriate for description of the neutrino spin evolution in 
electromagnetic fields.

We obtain \cite{LES99, hep-ph/9910476} the Lorentz invariant
generalization of eq.(\ref{A1}) for the case of $P$ invariance
violating theory demanding that the equation has to be linear over
spin vector $S_{\mu}$ and electromagnetic field $F_{\mu\nu}$.
Nonconservation of $P$ invariance implies existence of a preferred
direction in space in any reference frame.  The only choice of this
direction is given by vector ${\vec n}={\vec \beta}/\beta$.  Thus,
the Lorentz invariant generalization of eq.(\ref{A1}) can be obtained
by the substitution of the electromagnetic field tensor
$F_{\mu\nu}=(\vec E,\vec B)$ in the following way:  \begin{equation}
\begin{array}{c} F_{\mu\nu}\rightarrow F_{\mu\nu}+G_{\mu\nu},
\end{array}
\label{A2}
\end{equation}
where  the anti symmetric tensor $G_{\mu\nu}$ is constructed on the
base of the vector $\vec n$ in a way which is analogous to one by
which the electromagnetic tensor $F_{\mu\nu}$ is constructed on the
base of the polar vector $\vec E$ and axial vector $\vec B$:
\begin{equation}
\begin{array}{c}
G_{\mu\nu}=(\xi\vec n, \rho\vec n).
\end{array}
\label{A3}
\end{equation}
Here $\rho$ and $\xi$ are scalars. This substitution (\ref{A2})
in the case of the constant velocity, $\vec \beta=const$, implies
that the magnetic $\vec B$ and electric $\vec E$ fields are shifted
by the vectors $\rho\vec n$ and $\xi\vec n$,
respectively:
\begin{equation}
\begin{array}{c}
\vec B\rightarrow \vec B+\rho\vec n,
\vec E\rightarrow \vec E+\xi\vec n.
\end{array}
\label{A4}
\end{equation}
We finally
arrive \cite{LES99,hep-ph/9910476} to the following equation for the
three dimensional neutrino spin vector $\vec S$:  \begin{equation}
\begin{array}{c} {d\vec S \over dt}={2\mu \over \gamma} \Big[ {\vec
S} \times ({\vec B_0}+\rho {\vec n}) \Big]+{2\epsilon \over \gamma}
\Big[{\vec S} \times ({\vec E_0}+\xi {\vec n}) \Big].
\end{array}\label{BMT} \end{equation} The derivative in the left-hand
side of eq.(\ref{BMT}) is taken with respect to time $t$ in the
laboratory frame, whereas the values $\vec B_0$ and $\vec E_0$ are
the magnetic and electric fields in the neutrino rest frame
\begin{equation} \begin{array}{c} \vec B_0=\gamma\big(\vec B_{\perp}
+{1 \over \gamma} \vec B_{\parallel} + \sqrt{1-{1 \over
\gamma^2}}\big[{\vec E_{\perp} \times \vec n}\big]\big), \\ \vec
E_0=\gamma\big(\vec E_{\perp} +{1 \over \gamma} \vec E_{\parallel} -
\sqrt{1-{1 \over \gamma^2}}\big[{\vec B_{\perp} \times \vec
n}\big]\big), \end{array}\label{F0} \end{equation} where
\begin{equation} \begin{array}{c} \vec F_{\perp} =\vec F -\vec n
(\vec F \vec n), ~~ \vec F_{\parallel} =\vec n (\vec F \vec n), ~~
\vec F=~\vec B~or~\vec E, \end{array}\label{F} \end{equation} are the
transversal and longitudinal, in respect to the direction of the
neutrino motion components of magnetic and electric fields in the
laboratory frame.

 It should be noted here that the values of scalars $\rho$
and $\xi$ describe all kinds of interactions in which neutrinos
participate except the direct electromagnetic interaction of the
neutrino magnetic and electric moments with the external
electromagnetic field that are given by terms proportional to
${\vec B}_{0}$ and ${\vec E}_{0}$.
The explicit expressions for the values $\rho$ and $\xi$
depend on the considered model of the neutrino interaction.
Let us introduce the spin tensor \cite{LP99}
\begin{equation} \begin{array}{c} S={\vec \sigma}{\vec
S}, ~~ \vec \sigma =(\sigma_1, \sigma_2, \sigma_3),
\end{array}\label{S} \end{equation} that can be used for description
of the neutrino spin states (here $\sigma_i$ are the Pauli matrixes).
The behavior of $S=S(t)$ is given by the evolution operator~$U$,
\begin{equation} \begin{array}{c} S(t)=U S(t_{0}) U^{+}.
\end{array}\label{St} \end{equation}
For this operator one can get
the Schr$\ddot o$dinger type equation
\begin{equation}
\begin{array}{c}
i{d U \over d t}=HU
\end{array}\label{Sch}
\end{equation}
with the Hamiltonian given by,
\begin{equation}
\begin{array}{c}
H={\vec \sigma}{\vec R}, {\vec R}=-{\mu \over \gamma }({\vec
B}_{0}+\rho{\vec n}) -{\epsilon \over \gamma}({\vec E}_{0}+\xi{\vec
n}).
\end{array}\label{R}
\end{equation}
We consider the neutrino spin evolution for the case when
neutrino with mixing propagates through matter in presence of
electromagnetic field. In order to identify the values $\rho$ and
$\xi$, let us compare the Hamiltonian (\ref{R}) with the
corresponding part of the Hamiltonian usually used \cite{VVO} in
studies of the neutrino spin evolution in the transversal magnetic
field. In our notations the latter Hamiltonian generalized for
the case of various types of neutrino conversions with change of
helicity $\nu_{L}\leftrightarrow\nu_{R}$ can be written as
\begin{equation}
\begin{array}{c} H=({\vec \sigma}{\vec n})
\big({\Delta m^2A \over 4E} -{V \over 2}\big) -\mu{\vec
\sigma}\big({\vec B}_{\perp} +\big[{{\vec E_{\perp}} \times {\vec
n}}\big]\big)- \epsilon{\vec \sigma}\big({\vec E}_{\perp}
-\big[{{\vec B_{\perp}} \times {\vec n}}\big]\big).
\end{array}\label{VVO}
\end{equation}
The two parameters,
$A=A(\theta)$ being a function of vacuum mixing angle and
$V=V(n_{eff})$ being the difference of neutrino effective potentials
in matter depend on the nature of neutrino conversion processes in
question.  For specification of $A$ and $V$ for different types of
the neutrino conversions see, for example, in refs.\cite{LS,ALS}.
Then, taking into account $C$, $P$, and $T$
transformation properties of different terms of the Hamiltonian
we come from (\ref{R}), (\ref{F0})
and (\ref{VVO}) to the following
identification:
\begin{equation}
\begin{array}{c}
\mu\rho +\epsilon\xi~\rightarrow~\gamma \Big({V \over 2}-
{\Delta m^2A \over 4E}\Big).
\end{array}\label{rho}
\end{equation}

Finally we get the effective Hamiltonian that determines the evolution
of the
system $\nu=~(\nu_R,\nu_L)$ in presence of electromagnetic field with
given components $B_{\parallel, \perp}(t)$,
$E_{\parallel, \perp}(t)$ in the laboratory frame:
\begin{equation}
\begin{array}{c} H=({\vec \sigma}{\vec n}) \big({\Delta m^2A \over
4E} -{V \over 2}-{1 \over \gamma}(\mu B_{\parallel} + \epsilon
E_{\parallel})\big) -\mu{\vec \sigma}\big({\vec B}_{\perp}
+\big[{{\vec E_{\perp}} \times {\vec n}}\big]\big) \\ -\epsilon{\vec
\sigma}\big({\vec E}_{\perp} -\big[{{\vec B_{\perp}} \times {\vec
n}}\big]\big).
\end{array}\label{H} \end{equation}
In this expression for the Hamiltonian the terms proportional to
$1/\gamma^2$ and higher corrections in powers of $1/\gamma$ are
omitted. It should be noted that the difference of neutrino effective
potentials in matter, $V$, may contain also contributions 
\cite{ES,Nunokawa} from the medium polarization by the longitudinal 
magnetic field.

 As it follows from eq.(\ref{H}) the effective Hamiltonian
depends on the transversal $B_{\perp}$, $E_{\perp}$ and longitudinal
$B_{\parallel}$, $E_{\parallel}$ components
of the magnetic and electric fields.
Terms proportional to $B_\parallel$, $E_\parallel$ are suppressed by
a factor of $1/ \gamma \ll 1$ for the case of relativistic neutrinos.
However, for electromagnetic field configurations with strong enough
components $B_\parallel$ and $E_\parallel$ these terms can be
important.  In particular, as it will be shown below
the longitudinal component of the magnetic field could affect the
resonance condition in the neutrino oscillations $\nu_L
\leftrightarrow \nu_R$ (as well as in the helicity preserving neutrino
oscillations) through the direct interaction of the neutrino
magnetic moment with $B_\parallel$. This effect appears in addition
to the influence of the longitudinal magnetic field on
neutrino oscillations due to magnetic polarization of medium
\cite{ES, Nunokawa}.

Let us using the Hamiltonian (\ref{H}) consider the neutrino
spin precession in a field of electromagnetic wave with frequency
$\omega$. Here we suppose that the neutrino velocity is constant. We
denote by ${\vec e}_{3}$ the axis that is parallel with ${\vec n}$
and by $\phi$ the angle between ${\vec e}_{3}$ and the direction of
the wave propagation. For simplicity we shall neglect terms
proportional to the neutrino electric dipole moment $\epsilon$.  In
this case the magnetic field in the neutrino rest frame is given by
\begin{equation} \begin{array}{c} {\vec B_{0}}=~\gamma\big[B_1
(\cos{\phi}-\beta){\vec e}_{1} +B_2 (1-\beta\cos{\phi}){\vec e}_2 -{1
\over \gamma}B_1\sin{\phi}{\vec e}_{3}\big], \end{array}\label{B}
   \end{equation} where ${\vec e}_{1,2,3}$ are the unit orthogonal
vectors. For the electromagnetic wave of circular polarization
propagating in matter it is easy to get:  \begin{equation}
\begin{array}{c} B_1=B\cos{\psi}, ~B_2=B\sin{\psi},
\end{array}\label{B12} \end{equation} where $B$ is the amplitude of
the magnetic field in the laboratory frame and the phase of the wave
at the point where the neutrino is located at given time $t$ is
\begin{equation} \begin{array}{c} \psi=g\omega t(1-{\beta \over
\beta_{0}}\cos{\phi}).  \end{array}\label{B12_} \end{equation} The
phase depends on the wave speed $\beta_{0}$ in matter ($\beta_{0}\leq
1$). The values $g=\pm 1$ correspond to the two types of the circular
polarization of the wave.

The difference of the two terms, $(\cos\phi-\beta)$ and
$(1-\beta\cos\phi)$  of eq.(\ref{B_}) is proportional to
$1/\gamma^2$.  Neglecting of this difference, we obtain
\begin{equation}
\begin{array}{c}
{\vec R}=\Big(-{V \over
2} +{\Delta m^2A \over 4E} +{1 \over \gamma}\mu B_1
\sin{\phi}\big){\vec e}_{3} + \\
+\mu B(1-\beta\cos{\phi})({\vec
e}_{1}\cos{\psi} -{\vec e}_{2}\sin{\psi}).  \end{array}\label{Rwb}
\end{equation}

The exact solution of eq.(\ref{Sch}) with the Hamiltonian
determined by eq.(\ref{Rwb}) for the neutrino spin evolution in the
electromagnetic wave can be obtained in the case of $\sin\phi=0$ (
parallel or anti parallel propagation of the electromagnetic wave in
respect to the neutrino momentum). Moreover, it is possible to show
that for any direction of the wave propagation ($\sin\phi\not=0$)
the presence of the term $\gamma^{-1}B_{1}\sin\phi$ leads
to non sufficient changes of the type of the solution because
$B_1$ according to the definition (\ref{B12}) is an oscillating
function of time and there is also a suppression by a factor of
$\gamma^{-1}$.  If we neglect this term the solution of
eq.(\ref{Sch}) for the evolution operator $U(t)$ can be written in
the form \begin{equation} \begin{array}{c} U(t)=U_{{\vec
e}_{3}}(\psi-\psi_{0})U_{\vec l}~(\chi-\chi_{0}),
\end{array}\label{7} \end{equation} where \begin{equation}
\begin{array}{c} U_{{\vec e}_3}(\psi)=exp(i{\sigma}_3{\psi \over 2}),
U_{\vec l}~(\chi)=exp(i{({\vec \sigma}{\vec l})\over l}{\chi \over
2}).  \end{array}\label{7_} \end{equation} The evolution operator
$U(t)$ is a combination of the operator which describes rotation on
the angle $\chi-\chi_{0}=2l(t-t_{0})$ around the axis ${\vec l}$, and
the rotation operator on the angle $\psi-\psi_{0}$ around the axis
${\vec e}_{3}$ (the initial conditions for some time $t_0$ are fixed
by the angles $\psi_0$ and $\chi_0$).  For the vector ${\vec l}$ we
get \begin{equation} \begin{array}{c} {\vec l}=\Big({V \over 2}
-{\Delta m^2A \over 4E} -{{\dot \psi} \over 2} \Big){\vec e}_{3} -\mu
   B(1-\beta\cos{\phi})({\vec e}_{1}\cos{\psi_{0}} -{\vec
   e}_{2}\sin{\psi_{0}}).
   \end{array}\label{l} \end{equation}
From the exact expression (\ref{l}) for vector ${\vec l}$  one can
straightforwardly get probabilities of conversions
$\nu_{L}\leftrightarrow\nu_{R}$ between different types of neutrinos
with change of helicity.  It can be seen from eq.(\ref{l}) that the
amplitude of probability of conversion ($\sin^2{\theta_{eff}}$)
between the two neutrino helicity states could became sufficient, i.e.
$\sin^2\theta_{eff}\sim 1$, when vector ${\vec l}$ is orthogonal or
nearly orthogonal to the axis ${\vec e}_{3}$. This happens when the
condition
\begin{equation} \begin{array}{c} \Big| {V \over 2}
   -{\Delta m^2A \over 4E} -{g\omega \over 2}(1-{\beta \over
\beta_{0}} \cos{\phi}) \Big| \ll\mu B(1-\beta\cos{\phi})
\end{array}\label{Res_} \end{equation}
is satisfied.
It follows that the probability
amplitude of conversion $\nu_{L}\leftrightarrow\nu_{R}$ in the
electromagnetic wave could get its maximum value
($\sin^2{\theta_{eff}}=1$) for any strength of the field $B$
(it is supposed that $\mu B(1-\beta\cos\phi)\not= 0$) when the
resonance condition is fulfilled:
\begin{equation} \begin{array}{c}
   {V \over 2} -{\Delta m^2A \over 4E} -{g\omega \over 2}
   (1-{\beta \over \beta_{0}}
   \cos{\phi})=0.
\end{array}\label{Res__}
\end{equation}

Equation (\ref{Res__})
represents the new type of the resonance conditions for
the neutrino conversion processes $\nu_{L}\leftrightarrow\nu_{R}$
under the influence of the
field of the electromagnetic wave specified by the amplitude of the
magnetic field $B$, the frequency $\omega$, the polarization $g=\pm 1$, the
direction of propagation (given by the angle $\phi$) in respect to the
neutrino momentum, and the speed of propagation in matter
$\beta_{0}\leq 1$.  As it follows from eq.(\ref{Res__}) for the
fixed values of $\cos{\phi}$, $\beta$, and $\beta_{0}$ the resonance
condition for the particular conversion process
$\nu_{L}\leftrightarrow\nu_{R}$ can be satisfied only for one of
the two possible wave polarizations.

If for the particular conversion process
$\nu_{L}\leftrightarrow\nu_{R}$ the resonance condition is not
satisfied then from the unequality (\ref{Res_}) we can get (in a way
similar to our analysis for the case of the constant (and twisting)
magnetic field \cite{LS94} - \cite{LSg}) the critical strength of the
magnetic field of the electromagnetic wave,\hfill

\begin{equation} \begin{array}{c}
   B_{cr}={1 \over \mu(1- \beta\cos\phi)}\Big|\Big({V \over 2} -{\Delta m^2A \over 4E}
   -{g\omega \over 2} (1-{\beta \over \beta_{0}}
   \cos{\phi})\Big)\Big|,
\end{array}\label{Res__1}
\end{equation}
which determines a lower bound of the magnetic field for which the
oscillation amplitude is close to unity (i.e., at least it is not less
then $1/2$).

Now let us consider the neutrino conversion
$\nu_{L}\leftrightarrow\nu_{R}$ in the case when in addition to
electromagnetic wave given by eqs.(\ref{B12}) and (\ref{B12_}), a
constant longitudinal magnetic field, ${\vec
B}_{\parallel}=(0,0,B_{\parallel})$ is superimposed. The
effective magnetic field in the neutrino rest frame is given by
\begin{equation}
\begin{array}{c} {\vec B_{0}}=~\gamma\big[B_1
(\cos{\phi}-\beta){\vec e}_{1} +B_2 (1-\beta\cos{\phi}){\vec e}_2 +{1
   \over \gamma}\big(B_\parallel-B_1 \sin{\phi}\big){\vec
   e}_{3}\big].
\end{array}\label{B_} \end{equation}
Applying the results of the previous consideration for the case when
${\vec B}_{0}$ is determined by eq.(\ref{B_}) we get
\begin{equation}
\begin{array}{c}
{\vec R}=\Big(-{V \over 2} +{\Delta m^2A \over 4E} -{\mu \over
\gamma}\big(B_\parallel-B_1 \sin{\phi}\big) \Big){\vec e}_{3}+ \\
+\mu B(1-\beta\cos{\phi})({\vec e}_{1}\cos{\psi}
-{\vec e}_{2}\sin{\psi}).
\end{array}
\label{Rwb_}
\end{equation}
The solution of eq.(\ref{Sch}) is
given by eqs.(\ref{7}),(\ref{7_}), where for vector ${\vec l}$ we
get,
   \begin{equation} \begin{array}{c}
   {\vec l}=\Big({V \over 2}
   -{\Delta m^2A \over 4E}
   -{{\dot \psi} \over 2} +{1 \over \gamma} \mu B_\parallel\Big){\vec e}_{3}
   -\mu B(1 -\beta\cos{\phi})({\vec e}_{1}\cos{\psi_{0}}
   -{\vec e}_{2}\sin{\psi_{0}}).
   \end{array}\label{l_}
   \end{equation}
Therefore, the amplitude of the probability of conversion between the
   two neutrino helicity states in presence of the electromagnetic wave
and longitudinal magnetic field could become sufficient when the
following condition is satisfied,
\begin{equation}
\begin{array}{c}
\Big| {V \over 2} -{\Delta m^2A \over 4E} -{{\dot \psi} \over 2} +
{1 \over \gamma}\mu B_\parallel\Big| \ll\mu B(1-\beta\cos{\phi}).
\end{array}
\label{Res_2}
\end{equation}
The corresponding resonance condition now is:
\begin{equation}
\begin{array}{c}
{V \over 2} -{\Delta m^2A \over 4E} -{{g\omega} \over
2}(1-{\beta \over \beta_{0}}\cos\phi) + {1 \over \gamma}\mu
B_\parallel=0.  \end{array}\label{Res_3} \end{equation} It follows
that the longitudinal component of the magnetic field $B_{\parallel}$
could affect the resonance condition in the neutrino oscillations
$\nu_L \leftrightarrow \nu_R$  through the direct interaction of the
neutrino magnetic moment with $B_{\parallel}$.  This modification of
the resonance condition under the influence of $B_{\parallel}$ exists
in addition to the indirect effect of $B_{\parallel}$ that
can arise due to the polarization of medium in longitudinal
magnetic field \cite{ALS,ES,Nunokawa}.
The latter effect gives contribution to the difference of neutrino effective
potentials in matter, $V(n_{eff})$.

Equations (\ref{Res_2}) and (\ref{Res_3}) bind together the properties of
neutrinos ($\mu$, $\Delta m^2$, $E$, $\theta$) and
medium ($V$), as well as the direction of propagation, $\phi$, and other
characteristics of the electromagnetic wave (the frequency $\omega$,
the polarization $g$, the speed in matter $\beta_0$,
the strength of the field $B$) and the strength of the superimposed
longitudinal magnetic
field $B_\parallel$. Using the condition (\ref{Res_3}) we predict the new
type of resonances in the neutrino oscillations $\nu_L \leftrightarrow \nu_R$
that can exist in presence of the combination of electromagnetic wave
and constant longitudinal magnetic field.

Finally, we should like to emphasize the role of the direct
interaction of the neutrino magnetic (transition) moment with
longitudinal component of magnetic field. Consider neutrino moving in 
the presence of constant magnetic field ${\vec B}$. As it follows 
from the derived Hamiltonian (\ref {H}) even in vacuum $(V=0)$ left 
handed and right handed neutrino states are not maximally mixed by 
the presence of a magnetic field ${\vec B}={\vec B}_{\perp} +{\vec 
B}_{\parallel}$ unless ${\vec B}_{\parallel}$ vanishes exactly.  

Let us also discuss possibility for the neutrino resonance condition 
to be realized in the electromagnetic wave under 
the influence of the superimposed longitudinal magnetic field, 
${\vec B}_{\parallel}$.  Suppose that the term $({V \over 2} -{\Delta 
m^2A \over 4E})$ can be neglected in eqs.(\ref{Res_2}) and 
(\ref{Res_3}).  Then the critical field strength of the 
electromagnetic wave is given by \begin{equation} \begin{array}{c} 
B_{cr}={1 \over \mu(1- \beta\cos\phi)}\Big| -{g\omega \over 2} 
(1-{\beta \over \beta_{0}} \cos{\phi}) +{1 \over \gamma}\mu 
B_{\parallel}\Big|.  \end{array}\label{Res__1_} \end{equation} In the 
case when the neutrino is propagated along the electromagnetic wave 
   ($\cos \phi=1$) and $g=sign\{B_{\parallel}\}$ the corresponding 
resonance condition can be written in the form \begin{equation} 
\begin{array}{c} \omega=4\gamma\mu B_{\parallel}, 
\end{array}\label{omega}
\end{equation}
(we also take $\beta_{0}=1$).
If the condition (\ref{omega}) is fulfilled then
$B_{cr}\rightarrow 0$. Thus we conclude that the left handed and 
right handed neutrino states are maximally mixed even for very low 
strength of the field of the electromagnetic wave. Let us choose the 
frequency of the electromagnetic wave to be equal to the frequency of 
the microwave background radiation, $\omega\sim 2,5\cdot 10^{-4} eV$.  
Then from eq.  (\ref{omega}) we get the following expression for 
$B_{\parallel}$, \begin{equation} \begin{array}{c} 
B_{\parallel}=2,5\cdot 10^{-10}\gamma^{-1}{\mu_{0} \over \mu}B_{*}, 
\end{array} \label{B_parallel} \end{equation} where $\mu_{0}={e \over 
2m_{e}}$ is the Bohr magneton and $B_{*}={m^{2}_{e} \over 
e}=4,41\cdot 10^{13} \ Gauss$.  If one choose $m_{\nu}=1 eV$ and 
$E_{\nu}=1GeV$ it follows that $\gamma=10^{9}$ and for the value of 
the neutrino magnetic moment $\mu=\mu_{0}\cdot 10^{-10}$ from 
eq.(\ref{B_parallel}) it is possible to get estimation 
$B_{\parallel}\sim 10^{5} \ Gauss$.

The presence of the term ${1 \over \gamma}\mu B_{\parallel}$
(that describes the direct interaction of neutrinos with
${\vec B}_{\parallel}$) in the diagonal elements of the corresponding
Hamiltonians for neutrino conversions will also shift the resonance
condition in the case of neutrino oscillations without change of
helicity. These phenomena may contribute to the mechanisms of the
neutron star motion proposed previously (see refs.  \cite{KS, ALS}).
The effects discussed above can have important consequences for
neutrino oscillations in the other astrophysical environment. This
issue will be considered in detail elsewhere \cite{ELobanovS}.

 In conclusion we argue that the effective Hamiltonian for neutrino
oscillations (\ref{H}) can be used for description of neutrino
oscillations under the influence of an arbitrary electromagnetic
fields given by their components in the laboratory frame.

 We should like to thank Samoil Bilenky, Angelo Della Selva and
Lev Okun for helpful discussions.

\end{document}